\documentclass[pre,aps,showpacs,twocolumn,superscriptaddress,10pt,a4paper]{revtex4}
\usepackage{graphicx}

\begin{document}

\title{Quantum chaotic patterns in the E$\otimes(b_1+b_2)$ Jahn-Teller model }

\author{Eva Majern\'{\i}kov\'a}
\email{fyziemar@savba.sk}
\affiliation{Institute of Physics,
Slovak Academy of Sciences, D\'ubravsk\'a cesta 9, SK-84 511
Bratislava, Slovak Republic}
\affiliation{Department of
Theoretical Physics, Palack\'y University, T\v r. 17. Listopadu
50, CZ-77207 Olomouc, Czech Republic}

\author{S.Shpyrko}
\altaffiliation{On leave from the Institute of Nuclear Research,
Ukrainian Academy of Sciences, pr.Nauki 47 Kiev, Ukraine}
\affiliation{Department of Theoretical Physics, Palack\'y
University, T\v r. 17. Listopadu 50, CZ-77207 Olomouc, Czech
Republic}

\received{22 October 2005} \revised{23 March 2006} \published{24
May 2006}

\begin{abstract}
We study statistical properties of excited levels of the
E$\otimes(b_1+b_2)$ Jahn-Teller model. The multitude of avoided
crossings of energy levels is generally claimed to be a testimony
of quantum chaos. We found that apart from two limiting cases
($E\otimes e$ and Holstein model) the distribution of
nearest-neighbor spacings is rather stable as to the change of
parameters and different from the Wigner one. This limiting
distribution assumably shows scaling $\sim\sqrt{S}$ at small $S$
and resembles the semi-Poisson law $P(S)= 4S \exp (-2 S)$ at
$S\geq 1$. The latter is believed to be universal and
characteristic, e.g., at the transition between metal and
insulator phases.

\end{abstract}

\pacs{05.45.-a,31.30.-i,63.22.+m}

\maketitle

Phonon spectra  of two-level electron systems -- two phon\-on
E$\times$e Jahn-Teller (JT) model with rotation symmetry
\cite{Wagner:1992} and one-phonon exciton models
\cite{Kong:1990,Graham:1984,Eidson:1986,Esser:1994,Sonnek:1994,Herfort:2001}
show up remarkable features: multiple avoided level crossings
(MAC) and localized (``exotic'') excited states at certain quantum
numbers. These phenomena are typical for chaotic spectra (e.g., in
complex nuclei) usually associated with underlying non-integrable
Hamiltonians \cite{Eckhardt:1988,Nakamura:book:1993,Brody:1981}.

The reflection symmetric two-level Hamiltonians contain a hidden
nonlinearity due to the phonon assistance of the tunneling term
which reveals explicitly after the elimination of the electron
degrees of freedom
\cite{Wagner:1992,Shore:1973,Majernikova:2002,Wagner:1984,
Majernikova:2003}. Semiclassical approaches to the phonon dynamics
handle this nonlinearity in different ways. One or another
decoupling method results in losing different amounts of the
quantum information, and hence to controversial conclusions about
possible classical chaotic behavior \cite{Graham:1984}.

The adiabatic approach to the quantum E$\times$e
 JT model [rotation symmetric version with two vibron (boson) modes, one
 symmetric and the other antisymmetric against the reflection] applies
 in a limited range of validity for the strong
electron-phonon coupling
 \cite{Wagner:1992,Wagner:1984}.
Namely,  the rotational momentum $\hat{J}$ even in the ground
state ($|j|=\frac{1}{2}$) mediates the
 coupling between levels  and brings in the
 nonlinearity due to the reflection symmetry
 \cite{Wagner:1992,Majernikova:2002,Shore:1973,Wagner:1984,Majernikova:2003}.
Consequently, excited spectra especially at big $|j|$ are marked
by a high density of avoided level crossings \cite{Wagner:1992}.
The source of both MAC and localization is the interplay of the
quantum coupling of levels and nonlinearity. The asymmetry of the
interaction constants $\alpha \neq \beta$ in E$\otimes(b_1+b_2)$
JT model bears an additional source of the quantum
non-integrability.

\begin{figure}[b]
\includegraphics[scale=0.7]{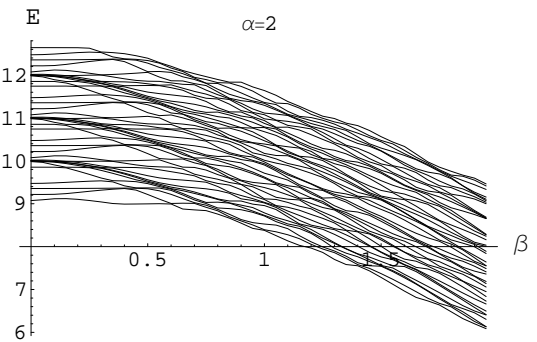}
\caption{ Complex behavior of energy levels of the nonsymmetric JT
model as function of $\beta$ - avoided crossings, crossings and
clusterings of levels.}
 \label{FigS6}
\end{figure}

Recently, Yamasaki {\it et al.}\cite{Nakamura:2003} first
investigated the E$\otimes$e JT model in terms of a search for
possible quantum chaotic patterns. The semiclassical decoupling
was performed according to the lines of the adiabatic approach.
The principal attention was paid however to an explicit nonlinear
term \cite{OBrien:1964} respecting the trigonal bulk symmetry. In
the previous work \cite{Majernikova:2005} we investigated the
quantum E$\otimes$e JT model numerically and analytically in
several limiting cases. We showed that an accurate account of the
hidden nonlinearity can lead to nontrivial patterns similar to
those produced in a system of two nonlinearly coupled oscillators.
Our numerical analysis showed the presence of the chaotic motion
domain at intermediate values of energy, which reflected in MAC in
the quantum spectrum. Appropriate patterns also revealed from the
statistical analysis of the nearest neighbor spacing distribution
(NNS) and the distribution of the ``level curvatures'' (second
derivatives of energies with respect to the coupling parameter
$\alpha$).

In the present paper we investigate the generalized model assuming
$\alpha\neq \beta$ [E$\otimes(b_1+b_2)$ model]. The local spinless
double degenerate electron level linearly coupled to two
intramolecular vibron (phonon) modes is described by the
Hamiltonian
\begin{equation}
\hat{H}=  (b_{1}^{\dag}b_{1} +b_{2}^{\dag}b_{2}+1 )I + \alpha
(b_{1}^{\dag}+b_{1})\sigma_{z}
 -\beta (b_{2}^{\dag}+b_{2})\sigma_{x},
 \label{1}
\end{equation}
where $\sigma_x$, $\sigma_z$ are $2\times 2$ Pauli matrices, $I$
is a unit matrix and the pseudospin notation refers to the
two-level electron system. The operators  $b_i$, $b_i^{\dag}$
satisfy boson commutation rules $[b_i,b_j^{\dag}]=\delta_{ij}$.
The interaction term $\propto\alpha$ removes the electron
degeneracy and the term $\propto\beta$ mediates the
phonon-assisted tunnelling.

The Hamiltonian (\ref{1}) has SU(2) symmetry and commutes with the
 reflection (parity) operator
\begin{equation}
\hat{R}=R_{ph}\sigma_x, \quad R_{ph}= \exp (i\pi b_1^{\dag}b_1),
\label{2}
\end{equation}
with $R_{ph}\hat Q_1=-\hat Q_1 R_{ph}$, $\hat Q_i\equiv
b_i^{\dag}+ b_i$; thus the eigenstates of the problem are chosen
to have a definite parity $p=\pm 1$. The phase plane $(\alpha,
\beta)$ includes two limiting effectively one-parameter models of
higher symmetry: (i) the rotation symmetric limit $\alpha=\beta$,
i.e. $E\otimes e$ Jahn-Teller case investigated previously
\cite{Majernikova:2005} and (ii) the one-phonon (Holstein) model
of either $\beta= 0$ or $\alpha= 0$. In the latter cases the wave
functions are coherent (displaced) Fock states $\exp
(\gamma(b_i^{\dag}-b_i)) |n\rangle$.

\begin{figure}[t]
\includegraphics[scale=0.55, clip=false]{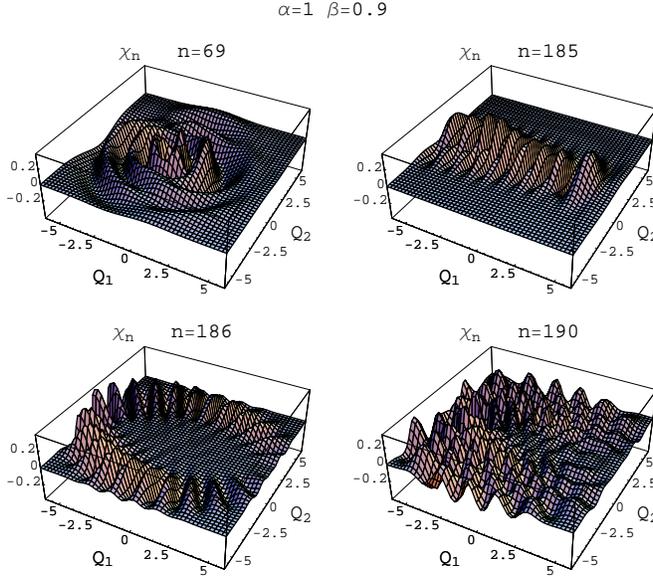}
\caption{(Color online) Examples of wave functions in the plane
$(Q_1 \otimes Q_2)$ for $\alpha=1$,$\beta=0.9$. Pronounced exotic
states close to either displaced Fock states or ``radial'' states
of the symmetric problem are shown. The extended (delocalized)
state $n=190$ is shown for comparison.}
 \label{FigW1}
\end{figure}

For arbitrary $(\alpha,\beta)$ the Hamiltonian (\ref{1}) can be
exactly diagonalized in the electron subspace using the
Fulton-Gouterman (FG) unitary operator \cite{Shore:1973}
\begin{equation}
 U= \frac{1}{\sqrt 2} \left ( \matrix{1\ , \ R_{ph}\cr  1\ ,
\ -R_{ph}}\right ) \label{38}
\end{equation}
In the radial coordinates $\hat Q_1\rightarrow r\cos\phi$ and
$\hat Q_2\rightarrow r\sin\phi$ the FG transformed Hamiltonian
(\ref{1}) for $p=+1$ is written as \cite{Majernikova:2005}
\begin{eqnarray}
\tilde{H}\equiv \hat{U} \hat{H} \hat{U}^{-1}=
-\frac{1}{2r}\frac{\partial}{\partial r} \left(
r\frac{\partial}{\partial r}\right) +\frac{1}{2}r^2
-\frac{1}{2 r^2} \frac{\partial^2}{\partial \phi^2}+\nonumber\\
 \sqrt{2}\alpha r \left(\cos\phi I-\sin\phi R_{ph}
  \sigma_z \right)
+ \sqrt{2}(\alpha-\beta )r \sin\phi R_{ph} \sigma_z \label{Ham}
\end{eqnarray}
[here the reflection (\ref{2}) acts as $R_{ph}
(r,\phi)f(r,\phi)=f(r,\pi-\phi)$]. Investigation of E$\otimes $e
case $\alpha=\beta$ in terms of rotational quantum numbers
[eigenvalues of conserved angular momentum $\hat{J}=i(b_1
b_2^+-b_1^+b_2)-1/2\sigma_y$] has a long history, dating back to
the paper in \cite{Long:1958}. The spectrum separates into
irreducible representations, each characterizing by the quantum
number $|j|=\frac{1}{2},\frac{3}{2}, \dots$. The whole matrix is
block-diagonal, and switching the term $\propto(\alpha-\beta)$
causes a complex interference of the levels in different blocks.
The complicated structure of the energy levels is exemplified in
Fig. 1 where we show 40 subsequent excited energy levels for
$\alpha=2$ and varying $\beta$. The level avoidings are
accompanied by level degeneracies (crossings). As in the symmetric
case, there is a number of excited wave functions showing
anomalous localization (Fig. 2) called "exotic states"
\cite{Wagner:1992}. One can recognize two distinct types of the
"exotic" states which are remnants of said limiting models with
higher symmetry: in Fig.2 the state $n=185$ reminds us of the
coherent state for $\beta=0$ while the states $n=69$ and $186$
with their markedly radial structure exemplify wave functions
typical for the case $\alpha=\beta$.

\begin{figure}[t]
\includegraphics[scale=0.6,clip=no]{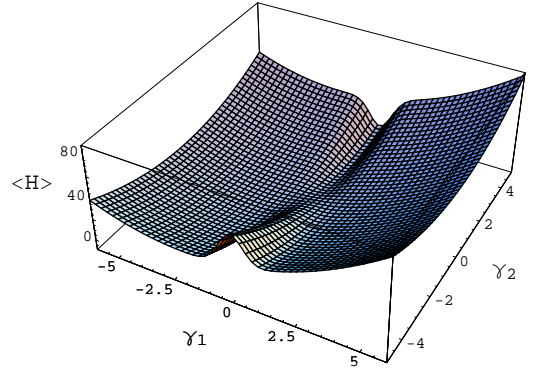}
\caption{(Color online) Effective potential built on coherent
states of both phonons. Its minima visualize complex interplay of
effective quantum oscillators.} \label{FigW3}
\end{figure}

Far from the rotation symmetry it is more convenient to perform
the transformation (\ref{38}) in the space $\hat Q_1 \times \hat
Q_2$:
 \begin{eqnarray}
 \tilde{H}_{FG} = \sum_{i=1,2}b_{i}^{\dag}b_{i}+1 +\alpha
(b_{1}^{\dag}+b_{1}) -p \beta (b_{2}^{\dag}+b_{2}) R_{ph}\,.
\label{39}
\end{eqnarray}
The elimination of electron degrees of freedom reveals the
nonlinearity hidden in the initial Hamiltonian (\ref{1}) [terms
with $R_{ph}$ in (\ref{Ham}) or (\ref{39}].
 Hamiltonian (\ref{39}) differs from that of exciton (dimer) by
the phonon-2 assistance in the tunneling term $\beta
(b_2^{\dag}+b_2) R_{ph}$ which accounts for Rabi oscillations by
the virtual emission and absorption of the phonon-$1$.
 These oscillations are essentially the origin of the
 nonlinearity of the reflection symmetric model and of its quantum
 nature. Namely, a consecutively classical version of the model would require to set
 $R_{ph}\equiv \pm 1$ dropping the nonlinear term. Thus
 the equivalence between the models (\ref{1}) and
(\ref{39}) is lost and the classical analog of the model (\ref{1})
with $\beta\neq 0$ is self-controversial.

 Averaging the diagonalized Hamiltonian (\ref{39}) over the trial
 wave function chosen as a combination of coherent states $\exp
[i\sum \limits_{k=1,2} R_{ph} (\gamma_k \hat P_k+\pi_k \hat
Q_k)]|0\rangle$ maps it onto a Husimi form \cite{Takahashi:1989}
for two oscillators nonlinearly coupled in variables
$\gamma_i,\pi_i$. The ``effective potential'' (Husimi
representation of the Hamiltonian operator)  built on the coherent
states in terms of the classical coordinates $\gamma_1, \gamma_2$
(Fig. 3)  was used for a variational treatment of the ground state
problem \cite{Majernikova:2003}. We showed that the parameter
space ($\alpha, \beta$) is separated onto two regions: the region
$\alpha>\beta$ of the dominating self-trapping or "heavy polaron"
with large $\gamma_1\sim - \alpha $ and small $\gamma_2\sim
\beta\exp (-2\gamma_1^2)$ and that ($\alpha<\beta$) of the
dominating tunneling between electron levels or "light polaron"
solution of small $\gamma_1 $ and large $\gamma_2$. Quantum
fluctuations cause mixing at the border between them and thus
broaden the transition region. The potential in Fig. 3 visualizes
the complex interplay of two oscillator potential wells resulting
in the emerging of the third very narrow local minimum (at
$\gamma_1\simeq 0$ and $\gamma_2>0$) responsible for the
appearance of the light polaron (for more detailed discussion on
this point see \cite{Majernikova:2002,Majernikova:2003}). This
heuristic visualization can be a guide for understanding phenomena
in the excited spectrum as well. Excited states will follow the
structure of either dominating self-trapping ($\alpha$) or
 tunneling ($\beta$) interactions.

\begin{figure}[t]
\includegraphics[scale=0.55]{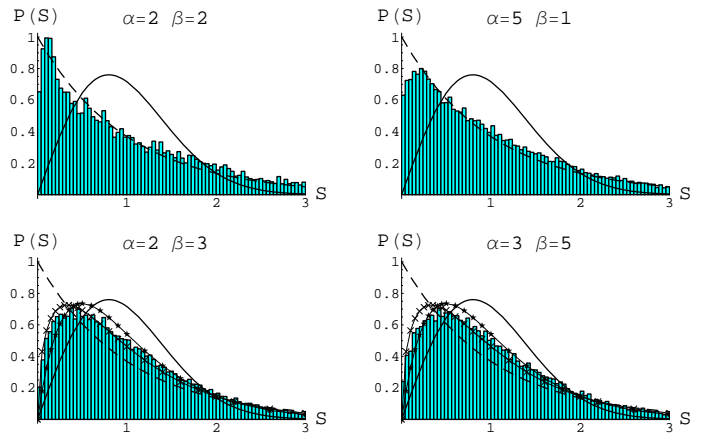}
\caption{(Color online) Nearest-neighbor distributions of levels
(unfolded and scaled to $\langle S \rangle = 1 $) for different
values of $\alpha$, $\beta$. The curves on the histograms
represent Poisson (long dashed), Wigner (full), semi-Poisson
(stars) and $\sqrt{S}$ (crosses) distributions.} \label{FigS7}
\end{figure}

The level spacing distributions are considered as leading
characteristics to distinguish between quantum integrability and
chaos, the latter being described by the Wigner surmise (WD)
$P_W(S)=(\pi/2)S\exp(-\pi S^2/4)$, and the former by the Poisson
statistics (PD) $P_P(S)=\exp(-S)$
\cite{Gaspard:1990,Mehta:1960,Brody:1981}. Since long ago the
Wigner surmise was conjectured as a limiting case for quantum
chaotic behavior and supported from the point of view of random
matrix theory (RMT; it is an exact conjecture for the 2$\times$2
RMT version, and a rather close approximation to the exactly
solvable case of random matrices of infinite dimensions
\cite{Mehta:1960}). The PD is associated with the superposition of
the multitude of uncorrelated levels: the falling exponential is a
limiting case of a big number of independent level sequences,
irrespectively to the level distributions inside each sequence
\cite{Rosenzweig:60}.  An adequate random-matrix model for our
case of broken symmetry requires separately treating block and
interblock elements \cite{Leitner:93,Leitner:94} yielding
complicated statistical predictions as to the resulting
superposition. Numerous interpolation formulas describing the
intermediate situations between the complete integrability and
chaos were considered \cite{Leitner:93,Brody:1981}.  A simple
interpolation based on the information theory considerations was
suggested
 \cite{Robnik:87} in
the form of the superposition of WD and PD: $P(S)\sim
(\pi/2)S\exp(-\mu S -\nu S^2)$. The accommodation constants $\mu$
and $\nu$ had to be chosen to ensure the normalization with
$\langle S \rangle=1$ conditions {\it and} yielding some {\it a
priori} given variance $\sigma^2$, thus giving a one-parameter
interpolation because the variance monotonically changes from
$(4/\pi-1)$ (WD) to $1$ (PD). Later on the theoretical background
for this form of interpolation was enhanced on the base of the
stochastic reformulation of the level statistics problem
\cite{Hasegawa:88}. The semi-Poisson distribution (sPD)
$P_{sP}(S)=4S \exp(-2S)$  with the dispersion exactly $0.5$
pertains to this class. Recently it was introduced to mimic new
seemingly universal properties in certain classes of systems, in
particular, being characteristics of the ``critical quantum
chaos'' \cite{Evangelou:00}, therefore it is worthy to probe it as
third reference point.

\begin{figure}[t]
\includegraphics[scale=0.5]{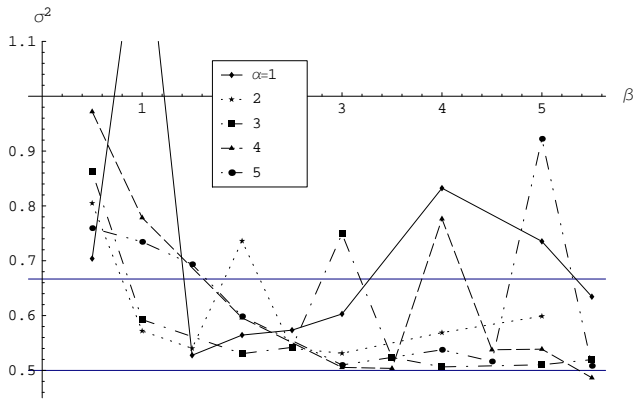}
\caption{Standard deviations $\sigma^2=\langle S^2 \rangle - 1$ of
the NNS distributions. Grid lines correspond to semi-Poisson (0.5)
and $\sqrt{S}$ ($\frac{2}{3}$) values.} \label{sigma}
\end{figure}

Figures 4 and 5 give the presentation of NNS statistics of JT
system with varying $\alpha$ and $\beta$. The level spacings were
unfolded according to the common procedure in order to exclude the
secular changes of level density and to ensure $\langle S
\rangle=1$. Figure 4 gives examples of NNS distributions for
sample values of $\alpha$ and $\beta$. Figure 5 presents the
standard deviations $\sigma^2\equiv \langle S^2\rangle - \langle S
\rangle^2$ as functions of $(\alpha, \beta)$. In the vicinity of
either $\alpha=\beta$, or $\alpha \gg \beta$, $\alpha \ll \beta$,
the distribution functions $P(S)$ follow rather Poisson statistics
(Fig. 4, first row), but small deviations of parameters from these
cases abruptly bring the distributions away from it. Figure 5,
however, reveals an interesting opposite universality for the
parameters far from the one-parametric cases (that is supposedly
in the most chaotic domain). The distribution functions seem to
tend to a well-defined limit, but this limiting case does not
resemble the Wigner surmise as one would expect. The standard
deviation of all curves has lower boundary equal to $0.5$ (in Fig.
5 it is markedly seen for the curves corresponding to $\alpha \ge
2$ in the domains approximately $0.4\alpha \leq \beta \leq
2.5\alpha$ but apart from $\beta\simeq \alpha$). Note also in
passing a remarkable mirror symmetry of interchange $\alpha
\leftrightarrow \beta$: the corresponding level statistics are
identical to a high degree of accuracy (the interchanged models
however are {\it not} equivalent, and, for example, the
wavefunctions of the ground state for $\alpha>\beta$ and
$\alpha<\beta$ differ essentially
\cite{Majernikova:2002,Majernikova:2003}).

Thus, the concurrence of two phonon modes of the Hamiltonian and
the existence of two symmetry-changing limits bring into being a
new limiting distribution which can be considered as a "most
quantum chaotic" one for this system. We cannot conclude whether
this hypothetic distribution is universal in the sense that it can
encounter elsewhere. In the scope of the present paper we only try
to mimic it discovering its possible universal properties. Natural
suggestion is to compare it with the semi-Poisson distribution
sharing the same $\sigma^2=0.5$ (Fig.4, second row).
Quantitatively, a coefficient of deviation from universal limits
of $P_W(S)$ and $P_P(S)$ can be introduced in the form
\cite{Shklovskii:93}
$\eta=\int_0^{S_0}[P(S)-P_W(S)]dS/\int_0^{S_0}[P_P(S)-P_W(S)]dS$
which gives the weight of the distribution left to the point
$S_0\simeq 0.4729$ of the intersection of WD and PD and ranges
from 0 (WD) to 1 (PD). For sPD $\eta_{sp}\simeq 0.3858$. The
graphs of $\eta$ similar to that of Fig.5 again show the existence
of a lower boundary $\eta\simeq 4.7 > \eta_{sp}$. Hence, the sPD
turns out to be much a better fit than WD, especially at $S\geq
1$. A systematic shift of the mass of the distribution to the left
with respect to sPD is markedly observed in Fig.4. This shift is
accounted for the nonlinear level repulsion of the actual
distributions. At small $S\ll 1$ they scale as $P\sim S^\delta$.
 The
actual repulsion index (Brody parameter) $\delta$ for "chaotic
sets" of $(\alpha,\beta)$ varies between $\sim 0.3$ and $0.5$,
meanwhile the class of WD and sPD assumes a linear repulsion
$P(S)\sim S$. Seeking for universal properties of a limiting
distribution we use the maximum value $\delta=0.5$ and suggest
another trial form $P_{sq}(S)\equiv (3\sqrt{3}/\sqrt{2\pi})
\sqrt{S} \exp(-3S/2)$. The coefficient in the exponent is chosen
to ensure $\langle S \rangle=1$ and to conform with the behavior
of sPD at large $S$. Samples in Fig.4 (second row) show that
$P_{sq}$-distribution reasonably fits actual distributions for
small $S$ in the chaotic domain, although the latter have a
tendency to shift slightly to the right. Therefore, the NNS
distributions far from the $E\otimes e$ and Holstein limits appear
to be confined between two suggested fitting formulas (note
corresponding grid lines in Fig.5). The standard $\chi^2$
reliability test however shows significant deviations between
actual distributions and both sPD and $P_{sq}$ indicating that
neither of the reference distributions is a good fit in the
strictly statistical sense.

 The sPD
 was recently suggested to describe a narrow intermediate region between
insulating and conducting regimes exemplified by the Anderson
localization model \cite{Evangelou:00}, the mentioned opposite
cases being described by correspondingly Poisson and Wigner
statistics. At present a plausible analytical support (in the
sense of RMT approaches) for this new distribution and for its
universal character is lacking. It was found numerically
\cite{Shklovskii:93} that the width of this intermediate domain
strongly depends on the length $L$ of the system (its number of
sites): for a system of infinite length one would get sPD in the
narrow region of the Anderson parameter around $W_{cr}$, meanwhile
for short lengths the width of the intermediate domain widens (in
\cite{Shklovskii:93} the lengths of the order of $\simeq 5-10$
were checked). Jahn-Teller systems from this point of view can be
considered as systems with $L=2$ (two electronic levels regarded
as pseudo-sites); thus it is to expect that the "transition
domain" for such a system is rather large and plain in the
parameter space, from whence there follows the applicability of
sPD for almost all reasonable values of parameters $\alpha,
\beta$. From the same point of view the systematic shift of the
distribution to the left of sPD indicates that the generalized JT
system is always closer to the "insulator" phase meaning the
dominance of "heavy" polarons localized on one electronic level
rather than the "light" ones. It is not surprising since heavy
polarons are associated with the broad well of the effective
potential (Fig.3) which has a higher density of states than the
narrow one responsible for light polarons. On the other hand, the
empirically suggested second reference distribution is of Brody
type \cite{Brody:1981} with the scaling $P\sim \sqrt{S}$ at small
$S$. The Brody parameter can be related \cite{Meyer:1984} to the
fraction of chaotic motion areas in the semiclassical picture. The
present work gives merely a sketch of these possible relations,
whose quantitative examination would be a challenge for the future
study.

 We acknowledge financial support from Project No. 202/06/0396 of the Grant
Agency of the Czech Republic. Partial support is acknowledged from
Project No. 2/6073/26 of the Grant Agency VEGA, Bratislava.

\end{document}